\documentclass{article}
\usepackage{amsmath}
\usepackage{amsfonts}
\usepackage{amssymb}
\usepackage{graphicx}

\input{tcilatex}

\begin{document}

\title{Updating Probabilities\thanks{%
Presented at MaxEnt 2006, the 26th International Workshop on Bayesian
Inference and Maximum Entropy Methods (July 8-13, 2006, Paris, France).}}
\author{Ariel Caticha and Adom Giffin \\
%EndAName
{\small Department of Physics, University at Albany-SUNY, }\\
{\small Albany, NY 12222, USA.}}
\date{}
\maketitle

\begin{abstract}
We show that Skilling's method of induction leads to a unique general theory
of inductive inference, the method of Maximum relative Entropy (ME). The
main tool for updating probabilities is the logarithmic relative entropy;
other entropies such as those of Renyi or Tsallis are ruled out. We also
show that Bayes updating is a special case of ME updating and thus, that the
two are completely compatible.
\end{abstract}

\section{Introduction}

The method of Maximum (relative) Entropy (ME) \cite%
{ShoreJohnson80,Skilling88,Caticha03} is designed for updating probabilities
when new information is given in the form of a constraint on the family of
allowed posteriors. This is in contrast with the older MaxEnt method \cite%
{Jaynes57} which was designed to assign rather than update probabilities.
The objective of this paper is to strengthen the ME method in two ways.

In \cite{Caticha03} the axioms that define the ME method have been distilled
down to three. In this work the justification of the method is improved by
considerably weakening the axiom that deals with independent subsystems. We
adopt a consistency axiom similar to that proposed by Shore and Johnson \cite%
{ShoreJohnson80}: When two systems are independent it should not matter
whether the inference procedure treats them separately or jointly. The merit
of such a consistency axiom is that it is very compelling. Nevertheless, the
mathematical implementation of the axiom has been criticized by Karbelkar 
\cite{Karbelkar86} and by Uffink \cite{Uffink95}. In their view it fails to
single out the usual logarithmic entropy as the unique tool for updating. It
merely restricts the form of the entropy to a one-dimensional continuum
labeled by a parameter $\eta $. The resulting $\eta $-entropies are
equivalent to those proposed by Renyi \cite{Renyi61} and by Tsallis \cite%
{Tsallis88} in the sense that they update probabilities in the same way.

The main result of this paper is to go beyond the insights of Karlbelkar and
Uffink, and show that the consistency axiom selects a unique, universal
value for the parameter $\eta $ and this value ($\eta =0$) corresponds to
the usual logarithmic entropy. The advantage of our approach is that it
shows precisely how it is that $\eta $-entropies with $\eta \neq 0$ are
ruled out as tools for updating.

Our second objective is mostly pedagogical. The preeminent updating method
is based on Bayes' rule and we want to discuss its relation with the ME
method. We start by drawing a distinction between Bayes' \emph{theorem},
which is a straightforward consequence of the product rule for
probabilities, and Bayes' \emph{rule}, which is the actual updating rule. We
show that Bayes' rule can be derived as a special case of the ME method, a
result that was first obtained by Williams \cite{Williams80,Diaconis82} long
before the logical status of the ME method had been sufficiently clarified.
The virtue of our derivation, which hinges on translating information in the
form of data into constraints that can be processed using ME, is that it is
particularly clear. It throws light on Bayes' rule and demonstrates its
complete compatibility with ME updating. A slight generalization of the same
ideas shows that Jeffrey's updating rule is also a special case of the ME
method.

\section{Entropy as a tool for updating probabilities}

Our objective is to devise a general method to update from a prior
distribution $q(x)$ to a posterior distribution $p(x)$ when new information
becomes available. By information, in its most general form, we mean a set
of constraints on the family of acceptable posterior distributions. \emph{%
Information is whatever constrains our beliefs}.

To carry out the update we proceed by ranking the allowed probability
distributions according to increasing \emph{preference}. This immediately
raises two questions: (a) how is the ranking implemented and (b) what makes
one distribution preferable over another? The answer to (a) is that any
useful ranking scheme must be transitive (if $P_{1}$ is better than $P_{2}$,
and $P_{2}$ is better than $P_{3}$, then $P_{1}$ is better than $P_{3}$),
and therefore it can be implemented by assigning a real number $S[P]$ to
each $P$ in such a way that if $P_{1}$ is preferred over $P_{2}$, then $%
S[P_{1}]>S[P_{2}]$. The preferred $P$ is that which maximizes the
\textquotedblleft entropy\textquotedblright\ $S[P]$. This explains why
entropies are real numbers and why they are meant to be maximized.

Question (b), the criterion for preference, is implicitly answered once the
functional form of the entropy $S[P]$ that defines the ranking scheme is
chosen. The basic strategy is inductive. We follow Skilling's method of
induction \cite{Skilling88}: (1) If an entropy $S[P]$ of universal
applicability exists, it must apply to special examples. (2) If in a certain
example the best distribution is known, then this knowledge constrains the
form of $S[P]$. Finally, (3) if enough examples are known, then $S[P]$ will
be completely determined. (Of course, the known examples might turn out to
be incompatible with each other, in which case there is no universal $S[P]$
that accommodates them all.)

It is perhaps worth emphasizing that in this approach entropy is a tool for
reasoning which requires no interpretation in terms of heat, multiplicities,
disorder, uncertainty, or amount of information. \emph{Entropy needs no
interpretation}. We do not need to know what it means, we only need to know
how to use it.

The known special examples, which are called the \textquotedblleft
axioms\textquotedblright\ of ME, reflect the conviction that what was
learned in the past is important and should not be easily ignored. The
chosen posterior distribution should coincide with the prior as closely as
possible and one should only update those aspects of one's beliefs for which
corrective new evidence has been supplied. The first two axioms are listed
below. (The motivation and detailed proofs are found in \cite{Caticha03}.)

\noindent \textbf{Axiom 1: Locality}. \emph{Local information has local
effects.}

\noindent When the new information does not refer to a domain $D$ of the
variable $x$ the conditional probabilities $p(x|D)$ need not be revised. The
consequence of the axiom is that non-overlapping domains of $x$ contribute
additively to the entropy: $S[P]=\int dx\,F(P(x),x)$ where $F$ is some
unknown function.

\noindent \textbf{Axiom 2: Coordinate invariance.} \emph{The ranking should
not depend on the system of coordinates. }

\noindent The coordinates that label the points $x$ are arbitrary; they
carry no information. The consequence of this axiom is that $S[P]=\int
dx\,m(x)\Phi (P(x)/m(x))$ involves coordinate invariants such as $dx\,m(x)$
and $P(x)/m(x)$, where the functions $m(x)$ (which is a density) and $\Phi $
are, at this point, still undetermined.

Next we make a second use of the locality axiom and allow domain $D$ to
extend over the whole space. Axiom 1 then asserts that \emph{when there is
no new information there is no reason to change one's mind. }When there are
no constraints the selected posterior distribution should coincide with the
prior distribution. This eliminates the arbitrariness in the density $m(x)$:
up to normalization $m(x)$ is the prior distribution, $m(x)\propto q(x)$.

In \cite{Caticha03} the remaining unknown function $\Phi $ was determined
using the following axiom:

\noindent \textbf{Old Axiom 3:\ Subsystem independence}. \emph{When a system
is composed of subsystems that are \textbf{believed} to be independent it
should not matter whether the inference procedure treats them separately or
jointly. }

\noindent Let us be very explicit about what this axiom means. Consider a
system composed of two subsystems which our prior evidence has led us to
believe are independent. This belief is reflected in the prior distribution:
if the subsystem priors are $q_{1}(x_{1})$ and $q_{2}(x_{2})$, then the
prior for the whole system is the product $q_{1}(x_{1})q_{2}(x_{2})$.
Further suppose that new information is acquired such that $q_{1}(x_{1})$ is
updated to $p_{1}(x_{1})$ and that $q_{2}(x_{2})$ is updated to $%
p_{2}(x_{2}) $. Nothing in this new information requires us to revise our
previous assessment of independence, therefore there is no need to change
our minds, and the function $\Phi $ must be such that the prior for the
whole system $q_{1}(x_{1})q_{2}(x_{2})$ should be updated to $%
p_{1}(x_{1})p_{2}(x_{2})$.

This idea is implemented as follows: First we treat the two subsystems
separately. Suppose that for subsystem $1$ maximizing 
\begin{equation}
S_{1}[P_{1},q_{1}]=\int dx_{1}\,q_{1}(x_{1})\Phi \left( \frac{P_{1}(x_{1})}{%
q_{1}(x_{1})}\right) ,  \label{S1[P1]}
\end{equation}%
subject to constraints $\mathcal{C}_{1}$ on the marginal distribution $%
P_{1}(x_{1})=\int dx_{2}\,P(x_{1},x_{2})$ selects the posterior $%
p_{1}(x_{1}) $. The constraints $\mathcal{C}_{1}$ could, for example,
include normalization, or they could involve the known expected value of a
function $f_{1}(x_{1})$, 
\begin{equation}
\int dx_{1}f_{1}(x_{1})P_{1}(x_{1})=\int
dx_{1}dx_{2}\,f_{1}(x_{1})P(x_{1},x_{2})=F_{1}~.  \label{C1}
\end{equation}%
Similarly, suppose that for subsystem $2$ maximizing the corresponding $%
S_{2}[P_{2},q_{2}]$ subject to constraints $\mathcal{C}_{2}$ on $%
P_{2}(x_{2})=\int dx_{1}\,P(x_{1},x_{2})$ selects the posterior $%
p_{2}(x_{2}) $.

Next we treat the subsystems jointly and maximize the joint entropy, 
\begin{equation}
S[P,q_{1}q_{2}]=\int dx_{1}dx_{2}\,q_{1}(x_{1})q_{2}(x_{2})\Phi \left( \frac{%
P(x_{1},x_{2})}{q_{1}(x_{1})q_{2}(x_{2})}\right) ,  \label{joint S[P]}
\end{equation}%
subject to the \emph{precisely the same constraints} on the joint
distribution $P$. The function $\Phi $ is determined by the requirement that
the selected posterior be $p_{1}p_{2}$. As shown in \cite{Caticha03} this
leads to the logarithmic form 
\begin{equation}
S[P,q]=-\int dx\,P(x)\log \frac{P(x)}{q(x)}~.  \label{S}
\end{equation}

\section{The new independence axiom}

Next we replace our old axiom 3 by an axiom which is more convincing axiom
because it is an explicit requirement of consistency.

\noindent \textbf{New Axiom 3:\ Consistency for independent subsystems}. 
\emph{When a system is composed of subsystems that are \textbf{known} to be
independent it should not matter whether the inference procedure treats them
separately or jointly.}

\noindent Again, we have to be very explicit about what this axiom means and
how it differs from the old one. When the subsystems are treated separately
the inference proceeds exactly as described before: for subsystem $1$
maximize the entropy $S_{1}[P_{1},q_{1}]$ subject to the constraints $%
\mathcal{C}_{1}$ to select a posterior $p_{1}$ and similarly for subsystem $%
2 $ to select $p_{2}$. The important difference is introduced when the
subsystems are treated jointly. Since we are only concerned with those
special examples where we \emph{know} that the subsystems are independent,
we are \emph{required} to search for the posterior within the restricted
family of joint distributions that take the form of a product $P=P_{1}P_{2}$%
; this is an \emph{additional} constraint over and above the original $%
\mathcal{C}_{1}$ and $\mathcal{C}_{2}$.

In the previous case we chose $\Phi $ so as to maintain independence because
there was no evidence against it. Here we impose independence by hand as an
additional constraint for the stronger reason that the subsystems are known
to be independent. At first sight it appears that the new axiom does not
place as stringent a restriction on the general form of $\Phi $: it would
seem that $\Phi $ has been relieved of its responsibility of enforcing
independence because it is up to us to impose it explicitly by hand.
However, as we shall see, the fact that we seek an entropy $S$ of \emph{%
general} applicability and that we require consistency for \emph{all possible%
} independent subsystems is sufficiently restrictive.

The new constraint $P=P_{1}P_{2}$ is easily implemented by direct
substitution. Instead of maximizing the joint entropy, $S[P,q_{1}q_{2}]$, we
now maximize 
\begin{equation}
S[P_{1}P_{2},q_{1}q_{2}]=\int dx_{1}dx_{2}\,q_{1}(x_{1})q_{2}(x_{2})\Phi
\left( \frac{P_{1}(x_{1})P_{2}(x_{2})}{q_{1}(x_{1})q_{2}(x_{2})}\right) ,
\label{joint S[P1P2]}
\end{equation}%
under independent variations $\delta P_{1}$ and $\delta P_{2}$ subject to
the same constraints $\mathcal{C}_{1}$ and $\mathcal{C}_{2}$ and we choose $%
\Phi $ by imposing that the updating leads to the posterior $%
p_{1}(x_{1})p_{2}(x_{2})$.

\subsection{Consistency for identical independent subsystems}

Here we show that applying the axiom to subsystems that happen to be
identical restricts the entropy functional to a member of the one-parameter
family given by 
\begin{equation}
S_{\eta }[P,q]=-\int dx\,P(x)\left( \frac{P(x)}{q(x)}\right) ^{\eta }\quad 
\text{for}\quad \eta \neq -1,0~.  \label{S eta}
\end{equation}%
Since entropies that differ by additive or multiplicative constants are
equivalent in that they induce the same ranking scheme, we could equally
well have written 
\begin{equation}
S_{\eta }[P,q]=\frac{1}{\eta (\eta +1)}\left( 1-\int dx\,P^{\eta +1}q^{-\eta
}\right) ~.  \label{S eta b}
\end{equation}%
This is convenient because the entropies for $\eta =0$ and $\eta =-1$ can be
obtained by taking the appropriate limits. For $\eta \rightarrow 0$ use $%
y^{\eta }=\exp \eta \log y\approx 1+\eta \log y$ to obtain the usual
logarithmic entropy, $S_{0}[P,q]=S[P,q]$ in eq.(\ref{S}). Similarly, for $%
\eta \rightarrow -1$ we get $S_{-1}[P,q]=S[q,P]$.

The proof below is based upon and extends a previous proof by Karbelkar \cite%
{Karbelkar86}. He showed that belonging to the family of $\eta $-entropies
is a sufficient condition to satisfy the consistency axiom for identical
systems and he conjectured but did not prove that this was perhaps also a
necessary condition. Although necessity was not essential to his argument it
is crucial for ours. We show below that for identical subsystems there are
no acceptable entropies outside this family.

\subsubsection*{Proof}

First we treat the subsystems separately. For subsystem $1$ we maximize the
entropy $S_{1}[P_{1},q_{1}]$ subject to normalization and the constraint $%
\mathcal{C}_{1}$ in eq.(\ref{C1}). Introduce Lagrange multipliers $\alpha
_{1}$ and $\lambda _{1}$, 
\begin{equation}
\delta \left[ S_{1}[P_{1},q_{1}]-\lambda _{1}\left( \int
dx_{1}f_{1}P_{1}-F_{1}\right) -\alpha _{1}\left( \int dx_{1}\,P_{1}-1\right) %
\right] =0,
\end{equation}%
which gives 
\begin{equation}
\Phi ^{\prime }\left( \frac{P_{1}(x_{1})}{q_{1}(x_{1})}\right) =\lambda
_{1}f_{1}(x_{1})+\alpha _{1}~,  \label{phi 1}
\end{equation}%
where the prime indicates a derivative with respect to the argument, $\Phi
^{\prime }(y)=d\Phi (y)/dy$. For subsystem $2$ we need only consider the
extreme situation where the constraints $\mathcal{C}_{2}$ determine the
posterior completely: $P_{2}(x_{2})=$ $p_{2}(x_{2})$.

Next we treat the subsystems jointly. The constraints $\mathcal{C}_{2}$ are
easily implemented by direct substitution and thus, we maximize the entropy $%
S[P_{1}p_{2},q_{1}q_{2}]$ by varying over $P_{1}$ subject to normalization
and the constraint $\mathcal{C}_{1}$ in eq.(\ref{C1}). Introduce Lagrange
multipliers $\alpha $ and $\lambda $, 
\begin{equation}
\delta \left[ S[P_{1}p_{2},q_{1}q_{2}]-\lambda \left( \int
dx_{1}f_{1}P_{1}-F_{1}\right) -\alpha \left( \int dx_{1}\,P_{1}-1\right) %
\right] =0,
\end{equation}%
which gives 
\begin{equation}
\int dx_{2}\,p_{2}\Phi ^{\prime }\left( \frac{P_{1}p_{2}}{q_{1}q_{2}}\right)
=\lambda \lbrack p_{2}]f_{1}(x_{1})+\alpha \lbrack p_{2}]~,  \label{phi 2}
\end{equation}%
where the multipliers $\lambda $ and $\alpha $ are independent of $x_{1}$
but could in principle be functionals of $p_{2}$.

The consistency condition that constrains the form of $\Phi $ is that if the
solution to eq.(\ref{phi 1}) is $p_{1}(x_{1})$ then the solution to eq.(\ref%
{phi 2}) must also be $p_{1}(x_{1})$, and this must be true irrespective of
the choice of $p_{2}(x_{2})$. Let us then consider a small change $%
p_{2}\rightarrow p_{2}+\delta p_{2}$ that preserves the normalization of $%
p_{2}$. First introduce a Lagrange multiplier $\alpha _{2}$ and rewrite eq.(%
\ref{phi 2}) as 
\begin{equation}
\int dx_{2}\,p_{2}\Phi ^{\prime }\left( \frac{p_{1}p_{2}}{q_{1}q_{2}}\right)
-\alpha _{2}\left[ \int dx_{2}\,p_{2}-1\right] =\lambda \lbrack
p_{2}]f_{1}(x_{1})+\alpha \lbrack p_{2}]~,
\end{equation}%
where we have replaced $P_{1}$ by the known solution $p_{1}$ and thereby
effectively transformed eqs.(\ref{phi 1}) and (\ref{phi 2}) into an equation
for $\Phi $. The $\delta p_{2}(x_{2})$ variation gives, 
\begin{equation}
\Phi ^{\prime }\left( \frac{p_{1}p_{2}}{q_{1}q_{2}}\right) +\frac{p_{1}p_{2}%
}{q_{1}q_{2}}\Phi ^{\prime \prime }\left( \frac{p_{1}p_{2}}{q_{1}q_{2}}%
\right) =\frac{\delta \lambda }{\delta p_{2}}f_{1}(x_{1})+\frac{\delta
\alpha }{\delta p_{2}}+\alpha _{2}~.
\end{equation}%
Next use eq.(\ref{phi 1}) to eliminate $f_{1}(x_{1})$, 
\begin{equation}
\Phi ^{\prime }\left( \frac{p_{1}p_{2}}{q_{1}q_{2}}\right) +\frac{p_{1}p_{2}%
}{q_{1}q_{2}}\Phi ^{\prime \prime }\left( \frac{p_{1}p_{2}}{q_{1}q_{2}}%
\right) =A[\frac{p_{2}}{q_{2}}]\Phi ^{\prime }\left( \frac{p_{1}}{q_{1}}%
\right) +B[\frac{p_{2}}{q_{2}}]~,  \label{phi 3}
\end{equation}%
where 
\begin{equation}
A[\frac{p_{2}}{q_{2}}]=\frac{1}{\lambda _{1}}\frac{\delta \lambda }{\delta
p_{2}}\quad \text{and}\quad B[\frac{p_{2}}{q_{2}}]=-\frac{\delta \lambda }{%
\delta p_{2}}\frac{\alpha _{1}}{\lambda _{1}}+\frac{\delta \alpha }{\delta
p_{2}}+\alpha _{2}~,
\end{equation}%
are at this point unknown functionals of $p_{2}/q_{2}$. Differentiating eq.(%
\ref{phi 3}) with respect to $x_{1}$ the $B$ term drops out and we get 
\begin{equation}
A[\frac{p_{2}}{q_{2}}]=\left[ \frac{d}{dx_{1}}\Phi ^{\prime }\left( \frac{%
p_{1}}{q_{1}}\right) \right] ^{-1}\frac{d}{dx_{1}}\left[ \Phi ^{\prime
}\left( \frac{p_{1}p_{2}}{q_{1}q_{2}}\right) +\frac{p_{1}p_{2}}{q_{1}q_{2}}%
\Phi ^{\prime \prime }\left( \frac{p_{1}p_{2}}{q_{1}q_{2}}\right) \right] ~,
\end{equation}%
which shows that $A$ is not a functional but a mere function of $p_{2}/q_{2}$%
. Substituting back into eq.(\ref{phi 3}) we see that the same is true for $%
B $. Therefore eq.(\ref{phi 3}) can be written as 
\begin{equation}
\Phi ^{\prime }\left( y_{1}y_{2}\right) +y_{1}y_{2}\Phi ^{\prime \prime
}\left( y_{1}y_{2}\right) =A(y_{2})\Phi ^{\prime }\left( y_{1}\right)
+B(y_{2})~,  \label{phi 4}
\end{equation}%
where $y_{1}=p_{1}/q_{1}$, $y_{2}=p_{2}/q_{2}$, and $A(y_{2})$, $B(y_{2})$
are unknown functions of $y_{2}$. If we specialize to identical subsystems
for which we can exchange the labels $1\leftrightarrow 2$, we get 
\begin{equation}
A(y_{2})\Phi ^{\prime }\left( y_{1}\right) +B(y_{2})=A(y_{1})\Phi ^{\prime
}\left( y_{2}\right) +B(y_{1})~.  \label{phi 5}
\end{equation}%
To find the unknown functions $A$ and $B$ differentiate with respect to $%
y_{2}$, 
\begin{equation}
A^{\prime }(y_{2})\Phi ^{\prime }\left( y_{1}\right) +B^{\prime
}(y_{2})=A(y_{1})\Phi ^{\prime \prime }\left( y_{2}\right)  \label{b}
\end{equation}%
and then with respect to $y_{1}$ to get 
\begin{equation}
\frac{A^{\prime }(y_{1})}{\Phi ^{\prime \prime }\left( y_{1}\right) }=\frac{%
A^{\prime }(y_{2})}{\Phi ^{\prime \prime }\left( y_{2}\right) }=a=\func{const%
}~.
\end{equation}%
Integrating, 
\begin{equation}
A(y_{1})=a\Phi ^{\prime }\left( y_{1}\right) +b~.
\end{equation}%
Substituting back into eq.(\ref{b}) and integrating gives 
\begin{equation}
B^{\prime }(y_{2})=b\Phi ^{\prime \prime }\left( y_{2}\right) \quad \text{and%
}\quad B(y_{2})=b\Phi ^{\prime }\left( y_{2}\right) +c~,
\end{equation}%
where $b$ and $c$ are constants. We can check that $A(y)$ and $B(y)$ are
indeed solutions of eq.(\ref{phi 5}). Substituting into eq.(\ref{phi 4})
gives 
\begin{equation}
\Phi ^{\prime }\left( y_{1}y_{2}\right) +y_{1}y_{2}\Phi ^{\prime \prime
}\left( y_{1}y_{2}\right) =a\Phi ^{\prime }\left( y_{1}\right) \Phi ^{\prime
}\left( y_{2}\right) +b\left[ \Phi ^{\prime }\left( y_{1}\right) +\Phi
^{\prime }\left( y_{2}\right) \right] +c~.  \label{phi 6}
\end{equation}%
This is a peculiar differential equation. We can think of it as one
differential equation for $\Phi ^{\prime }\left( y_{1}\right) $ for each
given constant value of $y_{2}$ but there is a complication in that the
various (constant) coefficients $\Phi ^{\prime }\left( y_{2}\right) $ are
themselves unknown. To solve for $\Phi $ choose a fixed value of $y_{2}$,
say $y_{2}=1$, 
\begin{equation}
y\Phi ^{\prime \prime }\left( y\right) -\eta \Phi ^{\prime }\left( y\right)
-\kappa =0~,  \label{phi 7}
\end{equation}%
where $\eta =a\Phi ^{\prime }\left( 1\right) +b-1$ and $\kappa =b\Phi
^{\prime }\left( 1\right) +c$. To eliminate the constant $\kappa $
differentiate with respect to $y$, 
\begin{equation}
y\Phi ^{\prime \prime \prime }+\left( 1-\eta \right) \Phi ^{\prime \prime
}=0~,  \label{phi 8}
\end{equation}%
which is a linear homogeneous equation and is easy to integrate. For a
generic value of $\eta $ the solution is 
\begin{equation}
\Phi ^{\prime \prime }(y)\propto y^{\eta -1}\Rightarrow \Phi ^{\prime
}(y)=\alpha y^{\eta }+\beta ~.
\end{equation}%
The constants $\alpha $ and $\beta $ are chosen so that this is a solution
of eq.(\ref{phi 6}) for all values of $y_{2}$ (and not just for $y_{2}=1$).
Substituting into eq.(\ref{phi 6}) and equating the coefficients of various
powers of $y_{1}y_{2}$, $y_{1}$, and $y_{2}$ gives three conditions on the
two constants $\alpha $ and $\beta $, 
\begin{equation}
\alpha (1+\eta )=a\alpha ^{2},\quad 0=a\alpha \beta +b\alpha ,\quad \beta
=a\beta ^{2}+2b\beta +c~.
\end{equation}%
The nontrivial ($\alpha \neq 0$) solutions are $\alpha =(1+\eta )/a$ and $%
\beta =-b/a$, while the third equation gives $c=b(1-b)/4a$. We conclude that
for generic values of $\eta $ the solution of eq.(\ref{phi 6}) is 
\begin{equation}
\Phi (y)=\frac{1}{a}y^{\eta +1}-\frac{b}{a}y+C~,  \label{sol a}
\end{equation}%
where $C$ is a new constant. Choosing $a=-\eta (\eta +1)$ and $b=1+Ca$ we
obtain eq.(\ref{S eta b}).

For the special values $\eta =0$ and $\eta =-1$ one can either first take
the limit of the differential eq.(\ref{phi 8}) and then find the relevant
solutions, or one can first solve the differential equation for general $%
\eta $ and then take the limit of the solution eq.(\ref{S eta b}) as
described earlier. Either way one obtains (up to additive and multiplicative
constants which have no effect on the ranking scheme) the entropies $%
S_{0}[P,q]=S[P,q]$ and $S_{-1}[P,q]=S[q,P]$.

\subsection{Consistency for non-identical subsystems}

Let us summarize our results so far. The goal is to update probabilities by
ranking the distributions according to an entropy $S$ that is of general
applicability. The functional form of the entropy $S$ has been constrained
down to a member of the one-dimensional family $S_{\eta }$. One might be
tempted to conclude (see \cite{Karbelkar86, Uffink95}) that there is no $S$
of universal applicability; that inferences about different systems ought to
be carried out with different $\eta $-entropies. But we have not yet
exhausted the full power of our new axiom 3.

To proceed further we ask: What is $\eta $? Is it a property of the
individual carrying out the inference or of the system under investigation?
The former makes no sense; we insist that the updating must be objective in
that different individuals with the same prior and the same information must
make the same inference. Therefore the \textquotedblleft inference
parameter\textquotedblright\ $\eta $ must be a characteristic of the system.

Consider two different systems characterized by $\eta _{1}$ and $\eta _{2}$.
Let us further suppose that these systems are independent (perhaps system $1$
is here on Earth while the other lives in a distant galaxy) so that they
fall under the jurisdiction of the new axiom 3; inferences about system $1$
are carried out with $S_{\eta _{1}}[P_{1},q_{1}]$ while inferences about
system $2$ require $S_{\eta _{2}}[P_{2},q_{2}]$. For the combined system we
are also required to use an $\eta $-entropy $S_{\eta
}[P_{1}P_{2},q_{1}q_{2}] $. The question is what $\eta $ do we choose that
will lead to consistent inferences whether we treat the systems separately
or jointly. The results of the previous section indicate that a joint
inference with $S_{\eta }[P_{1}P_{2},q_{1}q_{2}]$ is equivalent to separate
inferences with $S_{\eta }[P_{1},q_{1}]$ and $S_{\eta }[P_{2},q_{2}]$.
Therefore we must choose $\eta =\eta _{1}$ and also $\eta =\eta _{2}$ which
is possible only when $\eta _{1}=\eta _{2}$. But this is not all: any other
system whether here on Earth or elsewhere that happens to be independent of
the distant system $2$ must also be characterized by the same inference
parameter $\eta =\eta _{2}=\eta _{1}$ even if it is correlated with system $%
1 $. Thus all systems have the same $\eta $ whether they are independent or
not.

The power of a consistency argument resides in its universal applicability:
if a general expression for $S[P,q]$ exists then it must be of the form $%
S_{\eta }[P,q]$ where $\eta $ is a universal constant. The remaining problem
is to determine this universal $\eta $. One possibility is to determine $%
\eta $ experimentally: are there systems for which inferences based on a
known value of $\eta $ have repeatedly led to success? The answer is yes;
they are quite common.

The next step in our argument is provided by the work of Jaynes \cite%
{Jaynes57} who showed that statistical mechanics and thus thermodynamics are
theories of inference based on the value $\eta =0$. His method, called
MaxEnt, can be interpreted as the special case of the ME when one updates
from a uniform prior using the Gibbs-Shannon entropy. Thus, it is an
experimental fact without any known exceptions that inferences about \emph{%
all} physical, chemical and biological systems that are in thermal
equilibrium or close to it can be carried out by assuming that $\eta =0$.
Let us emphasize that this is not an obscure and rare example of purely
academic interest; these systems comprise essentially all of natural
science. (Included is every instance where it is useful to introduce a
notion of temperature.)

In conclusion: consistency for non-identical systems requires that $\eta $
be a universal constant and there is abundant experimental evidence for its
value being $\eta =0$. Other $\eta $-entropies may be useful for other
purposes but the logarithmic entropy $S[P,q]$ in eq.(\ref{S}) provides the
only consistent ranking criterion for updating probabilities that can claim
general applicability.

\section{Bayes updating}

The two preeminent updating methods are the ME method discussed above and
Bayes' rule. The choice between the two methods has traditionally been
dictated by the nature of the information being processed (either
constraints or observed data) but questions about their compatibility are
regularly raised. Our goal here is to show that these two updating
strategies are completely consistent with each other. Let us start by
drawing a distinction between Bayes' theorem and Bayes' rule.

\subsection{Bayes' theorem and Bayes' rule}

The goal here is to update our beliefs about the values of one or several
quantities $\theta \in \Theta $ on the basis of observed values of variables 
$x\in \mathcal{X}$ and of the known relation between them represented by a
specific model. The first important point to make is that attention must be
focused on the joint distribution $P_{\text{old}}(x,\theta )$. Indeed, being
a consequence of the product rule, Bayes' theorem requires that $P_{\text{old%
}}(x,\theta )$ be defined and that assertions such as \textquotedblleft $x$ 
\emph{and} $\theta $\textquotedblright\ be meaningful; the relevant space is
neither $\mathcal{X}$ nor $\Theta $ but the product $\mathcal{X}$ $\times
\,\Theta $. The label \textquotedblleft old\textquotedblright\ is important.
It has been attached to the joint distribution $P_{\text{old}}(x,\theta )$
because this distribution codifies our beliefs about $x$ and about $\theta $
before the information contained in the actual data has been processed. The
standard derivation of Bayes' theorem invokes the product rule, 
\begin{equation}
P_{\text{old}}(x,\theta )=P_{\text{old}}(x)P_{\text{old}}(\theta |x)=P_{%
\text{old}}(\theta )P_{\text{old}}(x|\theta )~,
\end{equation}%
so that%
\begin{equation}
P_{\text{old}}(\theta |x)=P_{\text{old}}(\theta )\frac{P_{\text{old}%
}(x|\theta )}{P_{\text{old}}(x)}~.  \tag{Bayes' theorem}
\end{equation}%
It is important to realize that at this point there has been no updating.
Our beliefs have not changed. All we have done is rewrite what we knew all
along in $P_{\text{old}}(x,\theta )$. Bayes' \emph{theorem} is an identity
that follows from requirements on how we should consistently assign degrees
of belief. Whether the justification of the product rule is sought through
Cox's consistency requirement and regraduation or through a Dutch book
betting coherence argument, the theorem is valid irrespective of whatever
data will be or has been collected. Our notation, with the label
\textquotedblleft old\textquotedblright\ throughout, makes this point
explicit.

The real updating from the old prior distribution $P_{\text{old}}(\theta )$
to a new posterior distribution $P_{\text{new}}(\theta )$ occurs when we
take into account the values of $x$ that have actually been observed, which
we will denote with a capital $X$. This requires a new assumption and the
natural choice is that the updated distribution $P_{\text{new}}(\theta )$ be
given by Bayes' \emph{rule}, 
\begin{equation}
P_{\text{new}}(\theta )=P_{\text{old}}(\theta |X)~.  \tag{Bayes' rule}
\end{equation}%
Combining Bayes' theorem with Bayes' rule leads to the standard equation for
Bayes updating, 
\begin{equation}
P_{\text{new}}(\theta )=P_{\text{old}}(\theta )\frac{P_{\text{old}}(X|\theta
)}{P_{\text{old}}(X)}~.
\end{equation}%
The assumption embodied in Bayes' rule is extremely reasonable: we maintain
those old beliefs about $\theta $ that are consistent with data values that
have turned out to be true. Data values that were not observed are discarded
because they are now known to be false.

This argument is indeed so compelling that it may seem unnecessary to seek
any further justification for the Bayes' rule assumption. However, we deal
here with such a basic algorithm for information processing -- it is
fundamental to all experimental science -- that even such a self-evident
assumption should be carefully examined and its compatibility with the ME
method should be verified.

\subsection{Bayes' rule from ME}

Our first concern when using the ME method to update from a prior to a
posterior distribution is to define the space in which the search for the
posterior will be conducted. We argued above that the relevant space is the
product $\mathcal{X}\times \Theta $. Therefore the selected posterior $P_{%
\text{new}}(x,\theta )$ is that which maximizes 
\begin{equation}
S[P,P_{\text{old}}]=-\tint dxd\theta ~P(x,\theta )\log \frac{P(x,\theta )}{%
P_{\text{old}}(x,\theta )}~
\end{equation}%
subject to the appropriate constraints.

Next, the information being processed, the observed data $X$, must be
expressed in the form of a constraint on the allowed posteriors. Clearly,
the family of posteriors that reflects the fact that $x$ is now known to be $%
X$ is such that 
\begin{equation}
P(x)=\tint d\theta ~P(x,\theta )=\delta (x-X)~.  \label{data constraint}
\end{equation}%
This amounts to an \emph{infinite} number of constraints: there is one
constraint on $P(x,\theta )$ for each value of the variable $x$ and each
constraint will require its own Lagrange multiplier $\lambda (x)$.
Furthermore, we impose the usual normalization constraint, 
\begin{equation}
\tint dxd\theta ~P(x,\theta )=1~.
\end{equation}

Maximize $S$ subject to these constraints, 
\begin{equation}
\delta \left\{ S+\tint dx\,\lambda (x)\left[ \tint d\theta ~P(x,\theta
)-\delta (x-X)\right] +\alpha \left[ \tint dxd\theta ~P(x,\theta )-1\right]
\right\} =0~,
\end{equation}%
and the selected posterior is 
\begin{equation}
P_{\text{new}}(x,\theta )=P_{\text{old}}(x,\theta )\,\frac{e^{\lambda (x)}}{Z%
}~,  \label{solution a}
\end{equation}%
where the normalization $Z$ is 
\begin{equation}
Z=\,e^{-\alpha +1}=\tint dxd\theta \,P_{\text{old}}(x,\theta )\,e^{\lambda
(x)}~,  \label{Z}
\end{equation}%
and the multipliers $\lambda (x)$ are determined from eq.(\ref{data
constraint}), 
\begin{equation}
\tint d\theta ~P_{\text{old}}(x,\theta )\frac{\,e^{\lambda (x)}}{Z}=P_{\text{%
old}}(x)\frac{\,e^{\lambda (x)}}{Z}=\delta (x-X)~.
\end{equation}%
Therefore, substituting $e^{\lambda (x)}$ back into eq.(\ref{solution a}), 
\begin{equation}
P_{\text{new}}(x,\theta )=\frac{P_{\text{old}}(x,\theta )\,\delta (x-X)}{P_{%
\text{old}}(x)}=\delta (x-X)P_{\text{old}}(\theta |x)~.
\end{equation}

The new marginal distribution for $\theta $ is%
\begin{equation}
P_{\text{new}}(\theta )=\tint dxP_{\text{new}}(x,\theta )=P_{\text{old}%
}(\theta |X)~,
\end{equation}%
which is Bayes' rule! Bayes updating is a special case of ME updating.

To summarize: the prior $P_{\text{old}}(x,\theta )=P_{\text{old}}(x)P_{\text{%
old}}(\theta |x)$ is updated to the posterior $P_{\text{new}}(x,\theta )=P_{%
\text{new}}(x)P_{\text{new}}(\theta |x)$ where $P_{\text{new}}(x)=\delta
(x-X)$ is fixed by the observed data while $P_{\text{new}}(\theta |x)=P_{%
\text{old}}(\theta |x)$ remains unchanged. Note that in accordance with the
philosophy that drives the ME method \emph{one only updates those aspects of
one's beliefs for which corrective new evidence has been supplied}.

The generalization to situations where there is some uncertainty about the
actual data is straightforward. In this case the marginal $P(x)$ in eq.(\ref%
{data constraint}) is not a $\delta $ function but a known distribution $%
P_{D}(x)$. The selected posterior $P_{\text{new}}(x,\theta )=\,P_{\text{new}%
}(x)P_{\text{new}}(\theta |x)$ is easily shown to be $P_{\text{new}%
}(x)=P_{D}(x)$ with $P_{\text{new}}(\theta |x)=P_{\text{old}}(\theta |x)$
remaining unchanged. This leads to Jeffrey's conditionalization rule, 
\begin{equation}
P_{\text{new}}(\theta )=\tint dx\,P_{\text{new}}(x,\theta )=\tint
dx\,P_{D}(x)P_{\text{old}}(\theta |x)~.
\end{equation}

\section{Conclusions}

We have shown that Skilling's method of induction has led to a unique
general theory of inductive inference, the ME method. The whole approach is
extremely conservative. First, the axioms merely instruct us what not to
update -- do not change your mind except when forced by new information.
Second, the validity of the method does not depend on any particular
interpretation of the notion of entropy -- entropy needs no interpretation.

Our derivation of the consequences of the new axiom show that when applied
to identical subsystems they restrict the entropy to a member of the $\eta $%
-entropy family. Its further application to non-identical systems shows that
consistency requires that $\eta $ be a universal constant which must take
the value $\eta =0$ in order to account for the empirical success of the
inference theory we know as statistical mechanics. Thus, the unique tool for
updating probabilities is the logarithmic relative entropy. Other entropies
with $\eta \neq 0$ or those of Renyi or Tsallis are ruled out; they may be
useful for other purposes but not for inference.

Finally we explored the compatibility of Bayes and ME updating. After
pointing out the distinction between Bayes' theorem and the Bayes' updating
rule, we showed that Bayes' rule is a special case of ME updating by
translating information in the form of data into constraints that can be
processed using ME.

\noindent \textbf{Acknowledgements:} We would like to acknowledge valuable
discussions with N. Caticha, R. Fischer, M. Grendar, K. Knuth, C. Rodr\'{\i}%
guez, and A. Solana-Ortega.

\end{document}